\documentclass[journal]{IEEEtranTIE}
\usepackage{amsmath,amsfonts}
\usepackage{newtxtext,newtxmath,amsmath}
\usepackage{algorithmic}
\usepackage{algorithm}
\usepackage{array}
\usepackage[caption=false,font=normalsize,labelfont=sf,textfont=sf]{subfig}
\usepackage{bbding}
\usepackage{textcomp}
\usepackage{stfloats}
\usepackage{url}
\usepackage{verbatim}
\usepackage{graphicx}
\usepackage{booktabs}
\usepackage{makecell}
\usepackage{amsmath}
\usepackage{multirow}
\usepackage{bm}
\usepackage{threeparttable}
\usepackage{soul}
\usepackage[backend=biber,style=ieee]{biblatex}

\usepackage[table,xcdraw]{xcolor}

\usepackage{tikz,xcolor,hyperref}
\hypersetup{hidelinks,
	colorlinks=true,
	allcolors=black,
	pdfstartview=Fit,
	breaklinks=true}
\definecolor{lime}{HTML}{A6CE39}
\DeclareRobustCommand{\orcidicon}{%
    \begin{tikzpicture}
    \draw[lime, fill=lime] (0,0) 
    circle [radius=0.16] 
    node[white] {{\fontfamily{qag}\selectfont \tiny ID}};    \draw[white, fill=white] (-0.0625,0.095) 
    circle [radius=0.007];    \end{tikzpicture}
    \hspace{-2mm}}
\foreach \x in {A, ..., Z}{%
    \expandafter\xdef\csname orcid\x\endcsname{\noexpand\href{https://orcid.org/\csname orcidauthor\x\endcsname}{\noexpand\orcidicon}}
    }

\begin{document}

\title{PE-TSFM: Self-Supervised Time-Series Learning for Generalizable Power Converter Health Monitoring under Unseen Conditions}

\author{Xinyuan Liao\orcidA{}, 
\IEEEmembership{Graduate Student Member, IEEE}, 
Xinyue Zhang\orcidB{}, \IEEEmembership{Member, IEEE},
Xing Wei\orcidC{}, \IEEEmembership{Member, IEEE},
Junwei Liu\orcidD{}, \IEEEmembership{Member, IEEE},
Shuai Zhao\orcidG{}, \IEEEmembership{Senior Member, IEEE},
Siqi Bu\orcidE{}, \IEEEmembership{Senior Member, IEEE},
and Yi Zhang\orcidH{}, \IEEEmembership{Senior Member, IEEE}
}

\markboth{Submitted to IEEE}%
{Shell \MakeLowercase{\textit{et al.}}: A Sample Article Using IEEEtran.cls for IEEE Journals}


\maketitle

\begin{abstract}
Data-driven health monitoring of power converters remains limited by poor generalization to unseen operating conditions. This work addresses this out-of-distribution (OOD) challenge by building a domain-specific time-series foundation model (PE-TSFM) that learns representations directly from large-scale unlabeled converter data. Unlike generic TSFMs trained on broad time-series datasets, the proposed PE-TSFM is pre-trained entirely on domain data, enabling it to learn the physical relationships unique to power electronics. To further tailor the model to this domain, we introduce a dual-attention mechanism that captures both temporal patterns and inter-channel dependencies. While generic TSFMs primarily model temporal dependencies, the added channel attention captures inter-sensor physical relationships essential for converter degradation analysis. A dataset containing 141 million unlabeled timestamps from an operating power converter is used for pre-training. Experiments show that PE-TSFM achieves 92\% accuracy under unseen operating conditions. In contrast, generic TSFMs achieve around 60\% and conventional time-series models achieve around 40\% accuracy. This result confirms the strong OOD generalization of the proposed PE-TSFM. Ablation studies further verify that the introduced channel attention mechanism significantly improves model performance. In addition, we conduct detailed studies on model scalability, hyperparameter sensitivity, and interpretability to provide a comprehensive understanding of the proposed approach.

\end{abstract}

\begin{IEEEkeywords}
Power electronics, condition monitoring, self-supervised learning, time-series foundation model.
\end{IEEEkeywords}

\definecolor{limegreen}{rgb}{0.2, 0.8, 0.2}
\definecolor{forestgreen}{rgb}{0.13, 0.55, 0.13}
\definecolor{greenhtml}{rgb}{0.0, 0.5, 0.0}

\section{Introduction}

\IEEEPARstart{C}{ondition} health monitoring of power converters is essential for various applications, such as traction converters, renewable energy, electric vehicles, and more \cite{alhmoud2018reliability, li2021high, li2022igbt}. Although numerous invasive solutions using additional sensors and dedicated measurement circuits have been proposed \cite{meng2025online}, industry practice still favors noninvasive methods that rely on existing converter sensors \cite{yu2025review}. Given the inherent complexity of converter degradation mechanisms, data-driven and AI-based approaches have recently gained significant attention in this field \cite{zhao2020overview}.

However, a major challenge of AI-based methods lies in their poor generalization when exposed to unseen or out-of-distribution (OOD) operating conditions. For example, models trained on laboratory data often fail to maintain accuracy when deployed in real traction converters operating under different mission profiles, such as varying load cycles, ambient temperatures, or control strategies \cite{distributionshift}. Similar performance degradation has been observed in other power electronic systems, such as photovoltaic inverters \cite{OODinverter} and motor drives \cite{ye2024ood}. These discrepancies between training and deployment conditions significantly limit the existing AI-based health monitoring solutions.

\begin{table}[t]
    \centering
    \caption{Evolution from machine learning to foundation models.}
    \renewcommand{\arraystretch}{1.25}
    \begin{tabular}{lll}
    \toprule
    & Typical methods & Learning objectives \\ \midrule
\begin{tabular}[c]{@{}l@{}}\vspace{-3pt}Machine Learning\\ (since 1990s)\end{tabular} & \begin{tabular}[c]{@{}l@{}}\vspace{-3pt}SVM,\\ Random forest\end{tabular} & \emph{\multirow{2}{*}{\begin{tabular}[c]{@{}l@{}}\vspace{-3pt}Learn data correlation\\ \vspace{-3pt}between input and output\\ for a specific task.\end{tabular}}} \\ 
\begin{tabular}[c]{@{}l@{}}\vspace{-3pt}Deep Learning\\ (since 2012)\end{tabular} & CNN, LSTM &  \\ \midrule
\begin{tabular}[c]{@{}l@{}}\vspace{-3pt}Foundation Model\\ (since 2020)\end{tabular} & \begin{tabular}[c]{@{}l@{}}\vspace{-3pt}GPT, BERT\\ \vspace{-3pt}(for natural\\ language, LLMs)\end{tabular} & \emph{\begin{tabular}[c]{@{}l@{}}\vspace{-3pt}Learn a transferable and \\ \vspace{-3pt}Invariant representations that\\ \vspace{-3pt}generalize across multiple \\ domains.\end{tabular}} \\ \bottomrule
\end{tabular}
    \label{tab:ai_models}
    \vspace{-5mm}
\end{table}

A fundamental reason is that the conventional task-specific AI models primarily learn data correlations from labeled data, as summarized in Table~\ref{tab:ai_models}. When the labeled dataset cannot cover all possible operating conditions, the trained model inevitably encounters a vast OOD space during deployment. This limitation is particularly severe in health monitoring, where labeled degradation or failure data are extremely scarce \cite{liu2024deep,yang2023failure}. While transfer learning has been explored to improve adaptability \cite{xia2021transferrable}, its effectiveness declines when the distribution shift between the training and OOD data is substantial. Moreover, transfer learning still requires extensive manual labeling efforts, which limit its scalability in industrial applications.

In recent years, the paradigm of data-driven methods has begun shifting from traditional deep learning to foundation models (FMs) \cite{bommasani2021opportunities}. These models are typically trained through self-supervised learning on large-scale unlabeled datasets, enabling them to learn transferable and invariant representations that generalize across multiple domains [see Table~\ref{tab:ai_models}]. This property has the potential to address the OOD challenges in converter health monitoring, as the model learns intrinsic temporal representations rather than overfitting to labeled conditions.

The overall FM pipeline is that first pre-training large-scale unlabeled data to learn generic feature representations, followed by fine-tuning on limited labeled data for task-specific adaptation. The concept of FMs first emerged in natural language processing, with landmark works such as BERT \cite{kenton2019bert} and ChatGPT \cite{achiam2023gpt}, also known as large language models (LLMs). Several recent studies in power electronics (PE) have applied pre-trained LLMs for design-related tasks \cite{zeid2024predicting,lin2024pe,pegpt2,pe3}, showing promising directions. However, PE health monitoring is fundamentally not a language-based domain, and there remains considerable debate over whether language-pretrained models can be effectively transferred to this domain-specific engineering problem.

Beyond language, researchers have explored FMs in time-series analysis \cite{guo2025swift}, such as Timer \cite{liu2024timer} and Mantis \cite{feofanov2025mantis}. These generic TSFMs are typically pre-trained on diverse time-series data across domains, such as finance, healthcare, meteorology, and sensor networks. They demonstrate that self-supervised pre-training can effectively capture universal temporal dependencies across datasets. Nevertheless, such models are not optimized for the physical dependencies and multimodal sensor signals inherent in PE systems. In PE, time-series data are the most representative information form, making the exploration of PE–specific TSFMs far more relevant.

From the above review, the noninvasive health monitoring in PE still faces two key challenges:
\begin{itemize}
    \item[1)] \emph{Limited labeled data and poor generalization}: existing data-driven methods rely heavily on scarce labeled datasets to learn task-specific correlations, making them highly sensitive to distribution shifts and unable to generalize under unseen or OOD operating conditions.
    \item[2)] \emph{Lack of domain-relevant FMs}: although recent FM research shows strong potential, most work in PE remains focused on applying LLMs or generic TSFMs. Domain-specific TSFMs, which align naturally with the multimodal sensor data of power converters, remain underexplored.
\end{itemize}

\begin{figure*}[t]
    \centering
    \includegraphics[width=160mm]{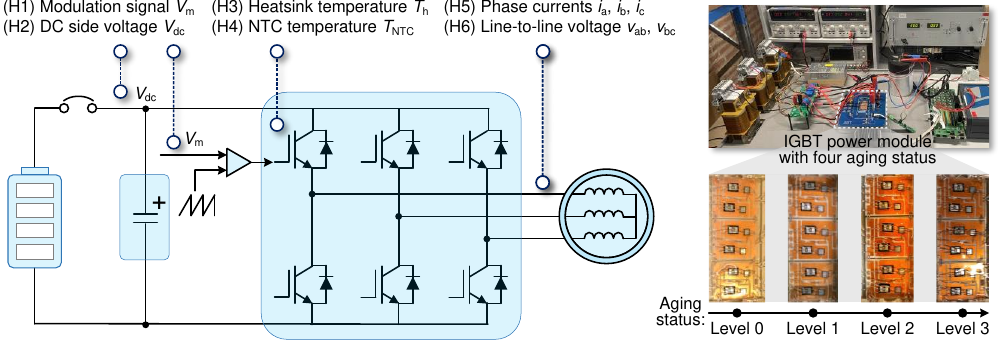}
    \caption{Experimental setup of a 5-kW traction converter used for health monitoring, focusing on the degradation of the power module. All signals (H1–H6) are captured noninvasively through the existing system infrastructure. The photos show the physical testbench and the same power module at four different aging levels (Level 0 – Level 3). Level 0 denotes a new device, while Level 3 approaches the end-of-life condition. The identical power module was stressed by the standard power cycling testing \cite{zhang2025power}, and each aging state was subsequently retested in the converter platform to collect data.}
    \vspace{-5mm}
    \label{fig:exp}
\end{figure*}

Building upon these insights, this paper proposes a PE-tailored TSFM (PE-TSFM) specifically designed for non-invasive health monitoring of power converters, focusing on the health status of power semiconductor devices. The main contributions are threefold:
\begin{itemize}
    \item[1)] \emph{Domain-specific foundation model}: Unlike prior works that rely on pretrained LLMs or generic TSFMs, the proposed PE-TSFM is fully developed using datasets collected from power converters, including both pre-training and fine-tuning. This domain-specific pipeline achieves superior performance compared with conventional data-driven methods and fine-tuned generic TSFMs.
    \item[2)] \emph{Tailored dual-attention mechanism}: Built upon the PatchTST and Transformer framework, the model introduces a dual-attention mechanism that combines temporal and channel attention. While generic TSFMs consider temporal dependencies only, the added channel attention models inter-sensor physical correlations in PE. Ablation results show that removing channel attention leads to a drop in accuracy of approximately 30\%, confirming its critical role in PE health monitoring.
    \item[3)] \emph{Self-supervised pre-training with unlabeled data for OOD generalization:} The PE-TSFM is trained on 143 million timestamps, with over 98\% unlabeled data used for self-supervised pre-training. Under unseen operating conditions, conventional data-driven methods achieve only 30–46\% accuracy, and generic TSFMs around 60\%, whereas the proposed PE-TSFM achieves up to 92\%, demonstrating its strong OOD generalization.
\end{itemize}

\section{Problem Definition and Dataset Description}

\subsection{Problem Formulation}

The health monitoring of power converters is formulated as a multivariate time-series classification problem. Let $x^i\in \mathbb{R}^{C\times L}$ denote observable time-series signals in power converters, where $C$ is the number of signal types (e.g., voltage, current, temperature) and $L$ is the length of the series set as 512, which is consistent with a common setting~\cite{Yuqietal-2023-PatchTST}. Let $y_i$ denote the ground-truth health label for the $i$-th observation. 

The goal of this task is to learn a nonlinear mapping function $f: \mathbb{R}^{C\times L}\xrightarrow{}\mathbb{R}^K$ where $K$ is the number of degradation levels, that transforms the input series into a vector of logits, which are then converted into a probability distribution via the softmax function. The predicted level is obtained by selecting the category with the highest logit:
\begin{equation}
    \hat{y}_i\xleftarrow{} \arg\max \left[ \text{softmax}(f( x_i) \right],
\end{equation}
where $\hat{y}_i$ is the predicted degradation level. This formulation enables end-to-end learning of discriminative temporal patterns from raw sensor data to health status classification.

\subsection{Experimental Platform and Defined OOD Conditions}

\begin{figure*}[t]
    \centering
    \includegraphics[width=165mm]{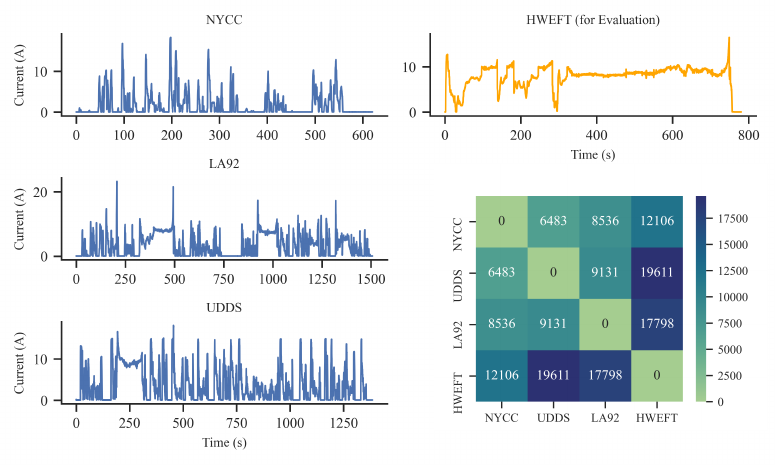}
    \vspace{-7mm}
    \caption{Euclidean norm of three-phase current $\|I\|_2$ under four standard driving cycles: NYCC, LA92, UDDS, and HWEFT. The heat map on the lower right shows the similarity matrix of different mission profiles, where smaller matrix values indicate higher similarity. The HWFET profile, which exhibits the largest deviation from the others, is reserved for OOD validation and excluded from model training. The similarity matrix is quantified by the dynamic time warping (DTW) algorithm \cite{salvador2007toward}.}
    \vspace{-5mm}
    \label{fig:mp}
\end{figure*}

To generate realistic operational data, experiments were conducted on a 5-kW traction converter platform designed for power module health monitoring, as shown in Fig.~\ref{fig:exp}. The converter includes an IGBT module and a DC-link capacitor operating under various dynamic mission profiles. The full setup and parameter configurations are detailed in \cite{xia2022impact}. All measurements were obtained through the existing controller interface and built-in sensors, ensuring that the data acquisition process remains noninvasive and does not alter converter operation.

Six categories of signals are collected: modulation reference $V_{\mathrm{m}}$, DC-link voltage $V_{\mathrm{dc}}$, phase currents $I_\text{a/b/c}$, line-to-line voltages $V_\text{ab}$ and $V_\text{bc}$, heatsink temperature $T_{\mathrm{h}}$, and power module-embedded NTC (negative–temperature–coefficient) temperature $T_{\mathrm{ntc}}$. The sampling rate is 10 kHz, enabling high-resolution monitoring of transient behaviors during dynamic mission profiles.

To capture device degradation, the same IGBT power module was subjected to standard power-cycling tests \cite{zhang2025power}, resulting in four representative aging levels: Level 0 (new), Level 1, Level 2, and Level 3 (severely aged approaching to end of life), as shown in Fig.~\ref{fig:exp}. The corresponding on-state voltages of the power module increased gradually from 2.97 V to 3.12 V, from Level 0 to Level 3. Each aged state is subsequently retested in the converter platform to collect its operational data.

Four standard driving cycles are used to emulate realistic converter operating conditions, as shown in Fig.~\ref{fig:mp}. The three in-distribution (ID) mission profiles: NYCC (New York City Cycle), LA92, and UDDS (Urban Dynamometer Driving Schedule). Although their diverse frequencies, load transients, and current waveforms, their similarity matrix values exhibit comparable statistical patterns. In contrast, the HWFET (Highway Fuel Economy Test) profile shows a substantially higher similarity matrix value (greater than 10,000) than all other profiles. Therefore, HWFET is reserved for OOD evaluation and excluded from model training, representing unseen mission profiles.


\begin{table}[t]
    \centering
    \caption{Dataset Configuration: Training on Massive Unlabeled Data and Testing under an Unseen Condition.}
    \renewcommand{\arraystretch}{1.25}
    \begin{tabular}{lllll}
    \toprule
         \multicolumn{2}{c}{Dataset} & \begin{tabular}[c]{@{}c@{}}\vspace{-4pt}Mission\\ profiles\end{tabular} & \begin{tabular}[l]{@{}l@{}}\vspace{-4pt}Total\\ timestamps\end{tabular} & Percentage \\
     \midrule
         \multirow{2}{*}{Training} & Unlabeled data & \multirow{2}{*}{Others} & 141 M & 98.2\% \\
        & Labeled data &  & 2.56 M & 1.8\% \\
    \midrule
        Testing & Unseen condition & HWFET & 30 M & -\\
    \bottomrule
    \end{tabular}
    \label{tab:dataset}
    \vspace{-5mm}
\end{table}

\subsection{Challenges in OOD Generalization}
Although four distinct aging levels were established through power-cycling tests, most of the dataset collected during converter operation remains unlabeled, as summarized in Table~\ref{tab:dataset}. Specifically, the power module operated under multiple mission profiles, generating over 143 million timestamps, of which only 2.56 million (1.8\%) were manually labeled with aging levels. The remaining 141 million (98.2\%) unlabeled samples reflect a realistic industrial scenario where vast operational data are available but lack degradation labels. Consequently, models must learn from scarce labeled data and generalize to unseen operating conditions.

For testing, an additional 30 million timestamps collected under the HWFET mission profile are reserved as unseen conditions and are not involved in model training (Table~\ref{tab:dataset}). Compared with the training missions, HWFET exhibits a distinct profile distribution (validated in Fig.~\ref{fig:mp}), providing a rigorous benchmark for evaluating the model’s OOD generalization capability. Consequently, the proposed learning framework must leverage scarce labeled data to achieve robust health monitoring performance under unseen conditions.

To quantitatively demonstrate the OOD challenge, four representative time-series models are evaluated using our constructed dataset, including the multilevel Wavelet Decomposition Network (mWDN) \cite{wang2018multilevel}, Fully Convolutional Network (FCN) \cite{wang2017time}, XceptionTime \cite{ismail2020inceptiontime}, and CNN-Transformer hybrid model (ConvTransPlus) \cite{foumani2024improving}. Two validation schemes are adopted. The in-distribution (ID) validation selects test samples from the same mission profiles used for training (NYCC, LA92, and UDDS). In contrast, the OOD validation employs the HWFET mission profile, which has never been seen during training and exhibits distinct statistical distributions.

As summarized in Table~\ref{tab:ood}, these models achieved only about 55\% accuracy in ID validation and experienced an additional 10–18\% performance drop under OOD conditions. These results highlight two key findings. First, with only 1.8\% labeled data, the models suffer from underfitting, indicating the difficulty of achieving high accuracy with limited supervision. Second, the significant degradation in OOD performance confirms that even established data-driven models struggle to maintain robustness under distribution shifts, underscoring the necessity for a new learning framework that can effectively generalize from unlabeled data.


\begin{table}[t]
    \centering
    \caption{In-distribution (ID) and out-of-distribution (OOD) accuracy of traditional data-driven methods}
    \begin{tabular}{lccc}
    \toprule
        \textbf{Model} & \textbf{ID (\%)} & \textbf{OOD (\%)} & \textbf{Difference (\%)}\\
        \midrule
         mWDN \cite{wang2018multilevel} & $55.77 \pm 1.25$ & $40.95 \pm 1.02$ &-14.82 \\
         FCN \cite{wang2017time}  & $52.37 \pm 1.24$ & $34.04 \pm 0.71$ & -18.33\\
         XceptionTime \cite{ismail2020inceptiontime}  & $57.23 \pm 0.87$ & $46.39 \pm 1.16$&-10.84 \\
         ConvTransPlus \cite{foumani2024improving}  &  $53.00 \pm 0.37$ &  $35.60 \pm 1.09$&-17.4 \\
        \bottomrule
        \end{tabular}
    \label{tab:ood}
    \vspace{-3mm}
\end{table}



\section{Proposed PE-TSFM Framework}

This section presents the proposed PE-TSFM. The framework introduces two main contributions. First, we develop a domain-specific TSFM trained entirely on power converter data, rather than adapting generic pre-trained TSFMs such as TiMER, Moment, or MANTIS, which rely on heterogeneous datasets and may fail to capture the unique PE time-series properties. Second, since the domain-specific TSFM contains both pre-training and fine-tuning stages, the model architecture can be flexibly redesigned. We propose a dual-attention mechanism that complements conventional temporal attention with an additional channel attention to capture inter-sensor physical correlations among the signals of power converters.

\subsection{Overview of TSFMs and Why They Can Learn Representations from Unlabeled Data}

Built on the dataset established in the previous section, this section presents the core contribution of this work: the PE-TSFM framework, a patch-based Transformer architecture with self-supervised pre-training. 

Unlike conventional task-specific models that rely on labeled data, TSFMs are designed to learn generalizable representations from large-scale unlabeled datasets. Through self-supervised pre-training, they discover intrinsic temporal dependencies by predicting, reconstructing, or contrasting masked portions of the input sequence. This process allows the model to encode the underlying structure and correlations of time-series data, capturing invariant temporal and cross-sensor relationships that persist across operating conditions. Once these representations are learned, they can be efficiently adapted to downstream tasks with only limited labeled data.

\begin{figure*}
    \centering
    \includegraphics[width=.9\linewidth]{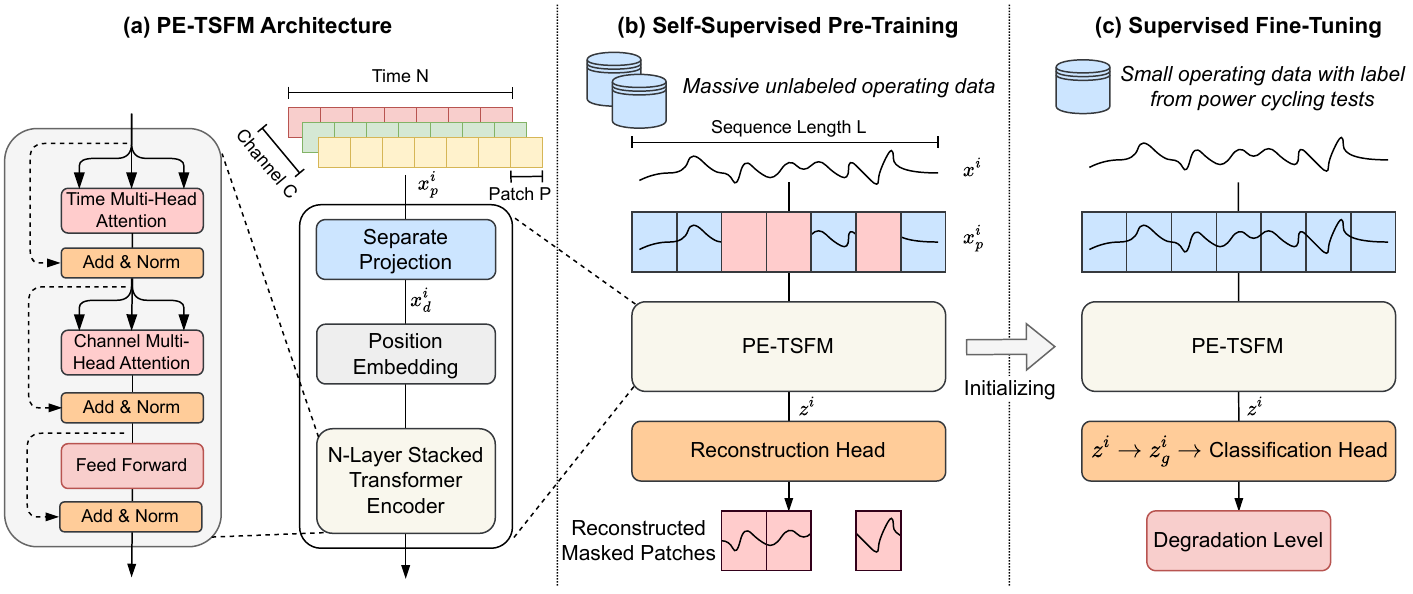}
    \caption{PE-TSFM architecture with the self-supervised pre-training and supervised fine-tuning pipelines for power module degradation monitoring.}
    \label{fig: backbone}
    \vspace{-3mm}
\end{figure*}

Fig.~\ref{fig: backbone}(b) conceptually illustrates this self-supervised representation learning process. The model first divides time-series data into small patches, randomly masks a portion of them, and encodes contextual dependencies using a Transformer encoder. The model is then trained to reconstruct or predict the masked segments. Through this iterative process, meaningful representations emerge without requiring manual labels, enabling robust generalization to unseen conditions.

\begin{table*}
    \centering
    \caption{A full comparison of mainstream time-series foundation models}
    \begin{tabular}{lcccccccl}\toprule
    Architecture & Pre-Train & Forecasting & Classification & Anomaly Detection & Imputation & Strength & Typical Model(s) \\
    \midrule
    GPT-Style &
    \makecell{Autoregressive  learning} &
    \Checkmark & \XSolidBrush & \XSolidBrush & \XSolidBrush &
    Forecasting &
    \makecell{Timer \cite{liu2024timer} \\ Moirai-MoE \cite{liu2025moiraimoe}} \\
    \midrule
    BERT-Style &
    \makecell{Masked  Reconstruction} &
    \Checkmark & \Checkmark & \Checkmark & \Checkmark &
    Understanding &
    \makecell{Moment \cite{goswami2024moment} \\ Moirai \cite{woo2024moirai} \\ \textbf{our PE-TSFM}} \\
    \midrule
    ViT-Style &
    \makecell{Contrastive  learning} &
    \XSolidBrush & \Checkmark & \Checkmark & \XSolidBrush &
    Discrimination &
    Mantis \cite{feofanov2025mantis} \\
    \bottomrule
    \end{tabular}
    \label{tab:tsfm}
\end{table*}

\subsection{Proposed PE-TSFM Architecture and Self-Supervised Pre-Training}

Among existing TSFM architectures, this work adopts the BERT-style masked reconstruction paradigm (Table~\ref{tab:tsfm}), which leverages full bidirectional attention to capture rich contextual dependencies within time-series data. This design enables the model to learn rich contextual representations that generalize across various conditions. Based on this paradigm, the proposed PE-TSFM framework utilizes the PatchTST backbone~\cite{Yuqietal-2023-PatchTST}, while introducing domain-specific enhancements tailored for degradation monitoring in power converters.

The overall architecture is shown in Figs.~\ref{fig: backbone}(a) and (b), comprising four major mechanisms: patching, positioning, dual-attention, and self-supervised learning. The key innovation lies in the proposed dual-attention mechanism. Traditional TSFMs pretrained on general datasets mainly rely on temporal attention, whereas signals in PE exhibit strong physical correlations across sensors. To address this, PE-TSFM introduces an additional channel attention to jointly capture not only temporal dependencies but also inter-sensor correlations. The following describes each component in detail.

\subsubsection{Patching}
The input multi-sensor sequence $x^i\in \mathbb{R}^{C\times L}$ is reshaped into a sequence of patches $x^i_p\in \mathbb{R}^{C\times N\times P}$, with $P$ denoting the patch size and $N$ the number of patches. Each patched sequence is subsequently projected into a higher-dimensional embedding $x^i_d\in \mathbb{R}^{C\times N\times d}$, where $d$ is the model hidden size. This process improves computational efficiency, enhances noise robustness, compresses redundant temporal patterns, and prepares the data for downstream attention-based encoding.

\subsubsection{Positional Embedding}
To capture temporal dynamics in power-electronics applications, it is essential to explicitly encode the position of each patch. This is achieved by adding positional encodings ($PosEn$) to the patch embeddings as $x^i_d+PosEn$. A widely used approach is the Sinusoidal Positional Encoding, introduced in the original Transformer model, which uses fixed, non-learnable functions defined as:
\begin{equation}
    \begin{aligned}
        &PosEn(pos, 2i)=sin(pos/10000^{2i/d}),\\
        &PosEn(pos, 2i+1)=cos(pos/10000^{2i/d}),
    \end{aligned}
\end{equation}
where $pos\in (0, N-1)$ denotes the patch's position, $i\in (0, d-1)$ denotes the index of embedding dimension.

\subsubsection{Dual Attention}
To capture degradation-relevant temporal–spatial patterns, PE-TSFM applies the scaled dot-product attention:
\begin{equation}
Attention(Q,K,V)=softmax(\frac{QK^{T}}{\sqrt{d}})V,
\end{equation}
where $Q,K,V\in\mathbb{R}^{d\times N}$ are derived from the patch embeddings $x_d^i$ through linear projections.

Multihead attention (MHA) extends this formulation to multiple representation subspaces, allowing the model to capture diverse dependencies. It is expressed as:
\begin{equation}
\begin{aligned}
    MHA(Q,K,V)=Concat(head_1,\cdots,head_n)W^o,\\
    head_i=Attention(Q_i,K_i,V_i).
\end{aligned}
\end{equation}

The proposed dual-attention mechanism jointly models two types of MHA: one for capturing temporal dependencies within each sensor (time MHA), and another for modeling inter-sensor relationships (channel MHA). These two attention mechanisms share the same parameters. Specifically, given an embedding $x_d^i \in \mathbb{R}^{C\times N\times d}$, the time MHA first computes attention scores along the temporal dimension $N$. The embedding is then transposed into $\mathbb{R}^{N\times C\times d}$, upon which the same attention layer (i.e., channel MHA) computes attention scores across the channel dimension $C$. This design captures channel-wise dependencies without introducing additional parameters, thereby improving computational efficiency. By stacking multiple encoder layers, the model generates a latent representation $z^i\in \mathbb{R}^{C\times N\times d}$ that encapsulates both time-wise and sensor-wise dynamics.

\subsubsection{Self-Supervised Learning Objective}
To enable representation learning from unlabeled time-series data, the self-supervised pre-training pipeline is shown in Fig.~\ref{fig: backbone}(b). First, a portion of the patches is masked using a constant mask value of $0$. Then, the remaining visible patches are processed by the Transformer encoder to generate a latent representation. This representation is passed through a linear reconstruction head to recover the masked patches, and the reconstruction error is used as the learning signal to update the model parameters. Formally, the pre-training objective is defined as
\begin{equation}
    \theta^* = \underset{\theta}{\arg\min} \frac{1}{|\mathcal{M}|} \sum_{j \in \mathcal{M}} \| f_\theta(p_{\text{masked}})_j - p_j \|^2
\end{equation}
where $\theta$ is the model parameter, $\mathcal{M}$ is the index set of masked patches, $f_\theta(p_{\text{masked}})_j$ is the reconstruction of the $j$-th masked patch, and $p_j$ is the $j$-th ground-truth patch. This objective drives the model to capture consistent temporal–spatial dependencies even in the absence of manual labels, enabling robust representation learning and improving generalization to unseen conditions.

\subsection{Supervised Fine-Tuning for Degradation Classification}

After the self-supervised pre-training stage, the PE-TSFM is fine-tuned using a small set of labeled degradation data to adapt the pretrained representations to the downstream classification task.
This stage focuses on transferring the general temporal–spatial features learned from massive unlabeled data to accurately distinguish degradation levels under practical operating conditions.

As shown in Fig.~\ref{fig: backbone}(c). The pretrained PE-TSFM encoder first extracts a latent representation for each input sequence. A global average pooling is used to extract a global feature vector $z^i_g\in \mathbb{R}^{C\times d}$ across the time dimension from the sequence representation $z^i$. Then, a linear classification head replaces the reconstruction head, flattening and mapping the global feature vector to degradation levels, and is trained using a small amount of labeled data (Table~\ref{tab:dataset}). This pre-training and fine-tuning two-stage approach enables effective adaptation to downstream health monitoring tasks with scarce labels.

\section{Experimental Results}


To validate the proposed PE-TSFM framework, the dataset in Table~\ref{tab:dataset} is utilized. Instance normalization \cite{kim2021reversible} is utilized to normalize all the data for better OOD generalization. All experiments are run on an AMD EPYC 7742 CPU and NVIDIA A100-SXM4-40GB GPU using PyTorch. The training details are provided in the Appendix.

\subsection{Validation for OOD Generalization}


\begin{table}[t]
    \centering
    \caption{Performance comparison with traditional data-driven methods and other foundation models under OOD conditions}
    \setlength{\tabcolsep}{4mm}
    \begin{tabular}{lcc}
    \toprule
        \textbf{Model} &  \textbf{Accuracy (\%)} & \textbf{F1-score (\%)} \\
        \midrule
        \multicolumn{3}{l}{\textbf{\textit{Traditional Data-Driven Methods}}} \\
         mWDN-1M (2018)&  $40.95 \pm 1.02$& $39.79 \pm 0.83$\\
         FCN-0.2M (2017)& $34.04 \pm 0.71$& $33.10 \pm 0.82$\\
         XceptionTime-0.4M (2020)&  $46.39 \pm 1.16$& $44.84 \pm 1.02$\\
         ConvTransPlus-26K (2024)&  $35.60 \pm 1.09$& $34.18 \pm 1.69$\\
         \midrule
         \multicolumn{3}{l}{\textbf{\textit{Generic TSFMs}}} \\
         Mantis-8M (2025)  & $57.33 \pm 6.03$ & $55.85 \pm 6.48$ \\
         Moment-35M8P (2024) & $66.64 \pm 2.74$ & $65.97 \pm 2.93$ \\
         \midrule
         \multicolumn{3}{l}{\textbf{\textit{Proposed PE-TSFM}}} \\
         PE-TSFM  & \bm{$92.73 \pm 0.82$} & \bm{$92.77 \pm 0.78$}  \\
        \bottomrule
    \end{tabular}
    \label{tab:compare}
    \vspace{-3mm}
\end{table}

The OOD generalization is a fundamental challenge for AI-based health monitoring methods. This study evaluates the proposed PE-TSFM under unseen operating conditions and compares it with both traditional data-driven methods and generic TSFMs. Specifically, we benchmark against four representative models: mWDN \cite{wang2018multilevel}, FCN \cite{wang2017time}, XceptionTime \cite{ismail2020inceptiontime}, and ConvTransPlus \cite{foumani2024improving}. These methods represent state-of-the-art architectures in wavelet methods, convolutional neural networks, and attention mechanisms.

The comparison results are summarized in Table~\ref{tab:compare}. The proposed PE-TSFM significantly outperforms all traditional data-driven methods in both accuracy and F1-score. It achieves an accuracy of 92.7\% and an F1-score of 92.7\%, far exceeding the best baseline (XceptionTime, 46.4\% accuracy). These results highlight the limitation of task-specific models, which cannot generalize across operating conditions due to the lack of unified representations. In contrast, PE-TSFM demonstrates superior self-learning and OOD generalization owing to its self-supervised pre-training and architecture co-design.

Moreover, we compare PE-TSFM with two generic TSFMs (i.e., Moment \cite{goswami2024moment} and Mantis \cite{feofanov2025mantis}), which are originally trained on broad cross-domain datasets (e.g., finance, energy, healthcare). For fair comparison, both models are fine-tuned on the same degradation classification dataset as PE-TSFM. The results indicate that PE-TSFM outperforms these models as well, demonstrating the importance of domain-specific pre-training. Despite their cross-domain capabilities, generic TSFMs fail to achieve excellent OOD generation in the health monitoring of power converters.

In summary, the proposed PE-TSFM achieves the best performance among all evaluated models, confirming that domain-specific TSFM provides strong OOD robustness for PE health monitoring.

\subsection{Ablation Analysis of Channel Attention and Pre-training}

\begin{figure}
    \centering
    \includegraphics[width=85mm]{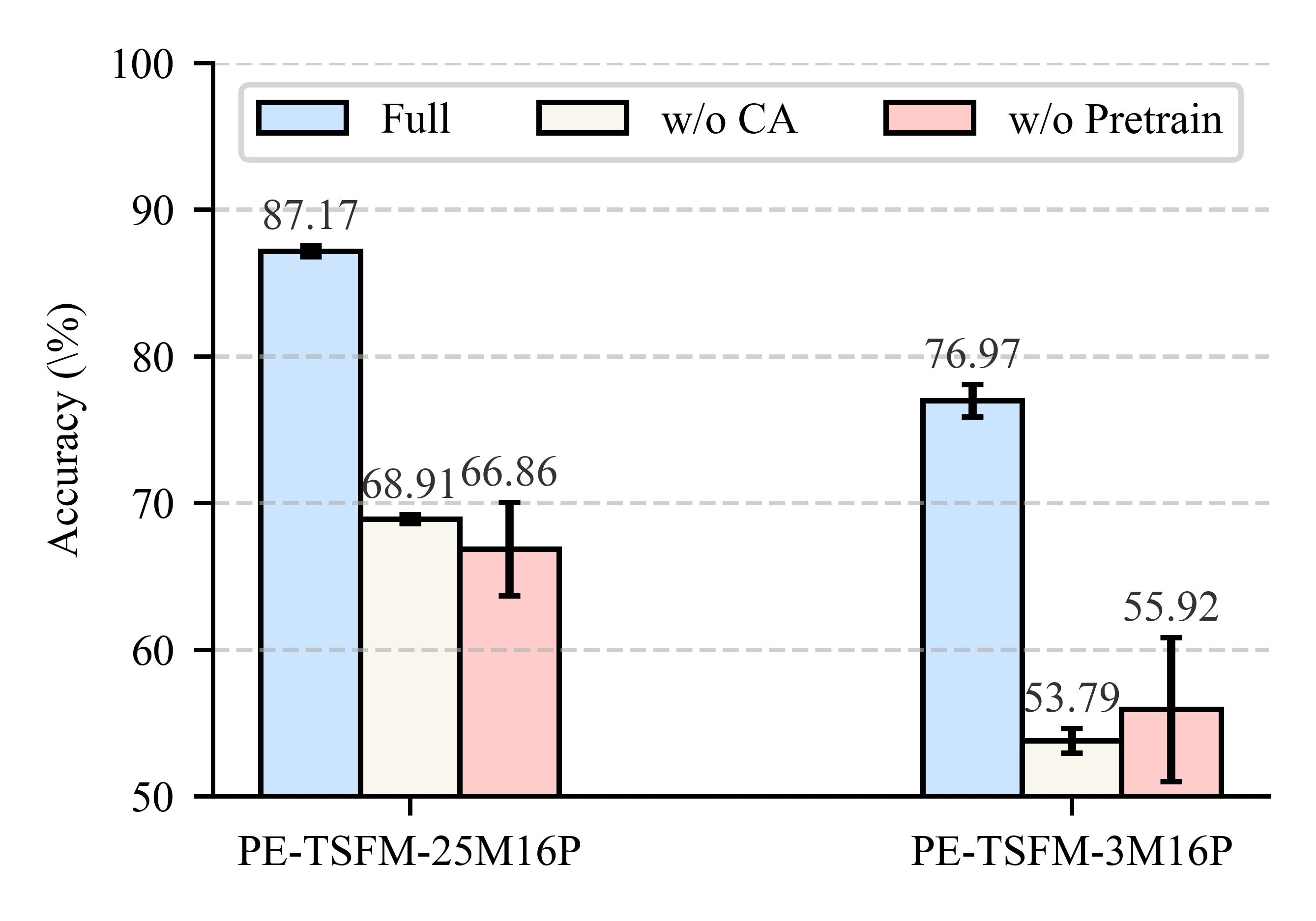}
    \vspace{-5mm}
    \caption{Classification performance, which was evaluated under three different random seeds, of full PE-TSFM and comparison models without channel attention and without pre-training.}
    \vspace{-5mm}
    \label{fig:ablation}
\end{figure}

This section analyzes the contribution of two core components in the proposed PE-TSFM: the channel attention (CA) mechanism and the self-supervised pre-training strategy in the domain-specific TSFM.

The CA mechanism is introduced to model inter-sensor dependencies, which are physically meaningful in power converters. For example, during the degradation of IGBT modules, the increase in on-state voltage and junction temperature arises from coupled electro-thermal effects, where electrical and thermal signals exhibit strong correlations. Conventional temporal-only attention fails to capture such cross-sensor relationships, leading to incomplete degradation representations. Meanwhile, the self-supervised pre-training stage enables the model to learn generalized temporal–spatial features from massive unlabeled data, providing robust initialization for downstream fine-tuning under scarce labeled conditions.

As shown in Fig.~\ref{fig:ablation}, removing either component leads to notable performance degradation. Excluding channel attention results in a clear accuracy drop, confirming that cross-sensor dependency modeling is essential for precise degradation classification. Eliminating the pre-training stage causes further accuracy loss and increased variance across runs, indicating unstable convergence and poor generalization.



\begin{table}[t]
    \centering
    \caption{Performance with different model parameters and patch sizes when batch size is 128}
    \begin{tabular}{lccc}
    \toprule
        \textbf{Model} & \textbf{Accuracy (\%)} & \textbf{F1-score (\%)} & \textbf{Memory (MB)}\\
        \midrule
         PE-TSFM-3M4P  & \bm{$92.73 \pm 0.82$} & \bm{$92.77 \pm 0.78$} &\textbf{12964} \\
         PE-TSFM-3M8P  & $83.75 \pm 3.09$ & $83.53 \pm 3.22$ &5866\\
         PE-TSFM-3M16P  & $76.97 \pm 1.11$ & $75.72 \pm 1.45$&3430 \\
         PE-TSFM-25M16P  &  $87.17 \pm 0.38$ &  $87.13 \pm 0.37$&10694 \\
        \bottomrule
        \end{tabular}
    \label{tab:perf}
    \vspace{-3mm}
\end{table}

\subsection{Impact of Model and Patch Size on Performance}
To evaluate the influence of model and patch size, four PE-TSFM variants were pre-trained with different configurations, as summarized in Table~\ref{tab:perf}. The baseline model, PE-TSFM-3M16P, contains about 3 million parameters, 4 layers, and a model dimension ($d_{\text{model}}$) of 256, using a patch size of 16. To further examine the effect of model parameter scaling, we pre-trained PE-TSFM-25M16P, which has about 25 million parameters, consisting of 8 layers and a $d_{\text{model}}$ of 512. In addition, to explore the impact of patch size on performance, we trained PE-TSFM-3M8P and PE-TSFM-3M4P, with patch sizes of 8 and 4, respectively. Each model has been tested three times under different random seeds.

\begin{figure}[!t]
    \centering
    \includegraphics[width=85mm]{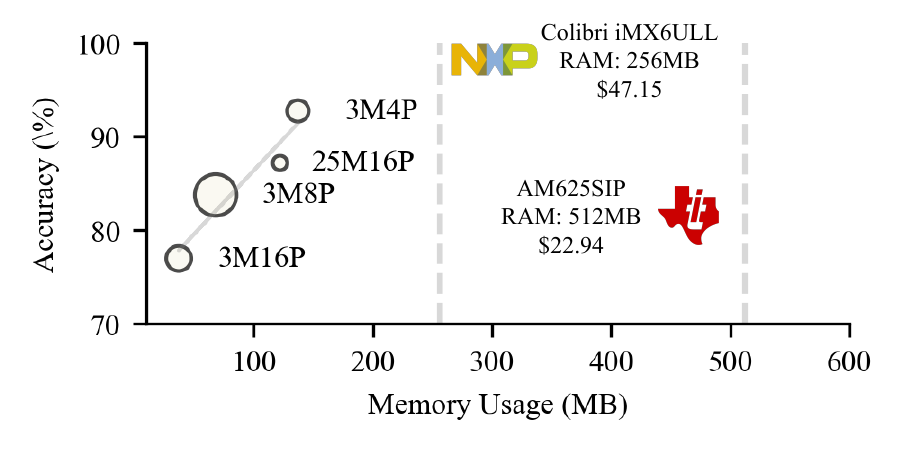}
    \vspace{-5mm}
    \caption{The relationship between GPU memory usage and model performance, where the bigger bubble denotes the model has a larger variance under different random seeds. Note that the GPU memory usage is measured during inference on 1 sample with FP32.}
    \vspace{-5mm}
    \label{fig:mem}
\end{figure}

As shown in Table~\ref{tab:perf} and Fig.~\ref{fig:mem}, both increasing model size and decreasing patch size improve accuracy. In fact, enlarging the model dimension increases the size of individual attention maps, whereas reducing the patch size raises the number of attention maps. Both strategies, therefore, contribute to higher model complexity. As shown in Fig.~\ref{fig:mem}, the relationship between model performance and GPU memory consumption is further examined, revealing an approximately linear trend. This finding provides a useful reference for future efforts in model scaling.

Moreover, while the pre-training of TSFM needs a high-performance GPU, the deployment of the most expensive PE-TSFM-3M4P only occupied 137 MB of GPU memory during inference on 1 sample with FP32, which makes it possible to be deployed on the edge-end compute device, such as NXP Colibri iMX6ULL with 256MB RAM and TI AM625SIP with 512MB RAM, for online degradation monitoring.

\subsection{Analysis of Degradation-Level Classification Accuracy}

Since PE-TSFM-3M8P exhibits large variance and lacks distinctive advantages, we focus our in-depth analysis on the other three models: PE-TSFM-3M16P, PE-TSFM-3M4P, and PE-TSFM-25M16P. 

\begin{figure*}
    \centering
    \includegraphics[width=165mm]{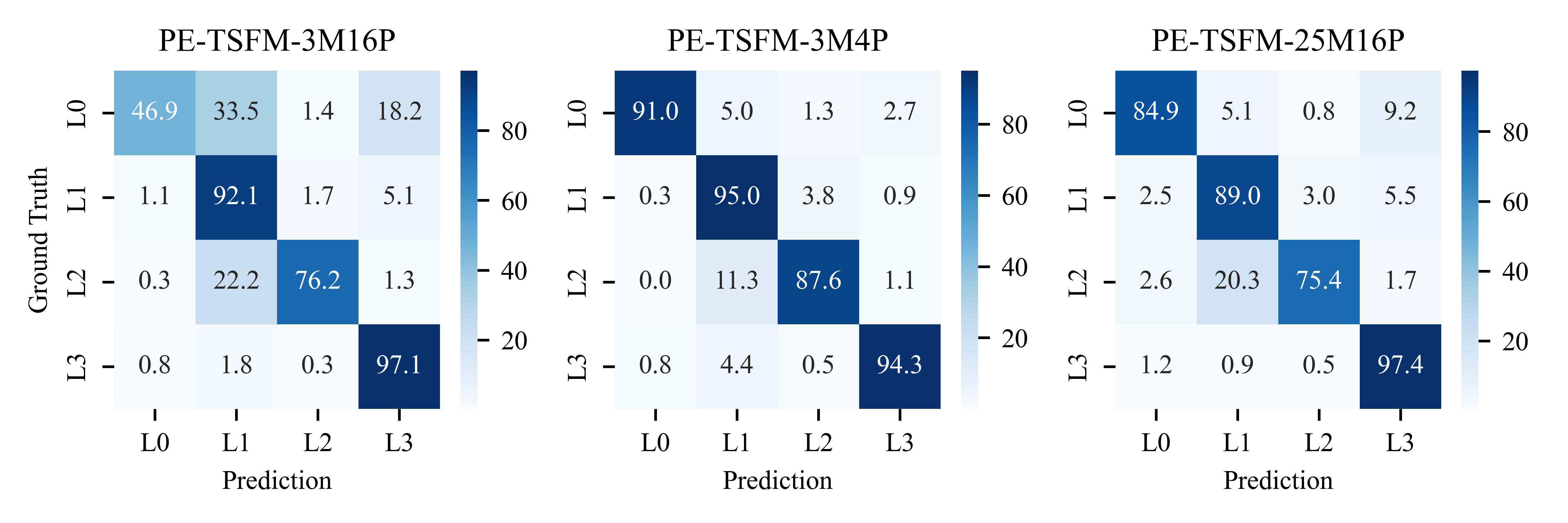}
    \vspace{-5mm}
    \caption{The normalized confusion matrices of PE-TSFM-3M16P, PE-TSFM-3M4P, and PE-TSFM-25M16P.}
    \vspace{-5mm}
    \label{fig:cm}
\end{figure*}

Fig.~\ref{fig:cm} presents the confusion matrices of the PE-TSFMs. The weakest baseline, PE-TSFM-3M16P, shows the most errors in degradation level 0, which reveals the inherent challenge of detecting early-stage degradation. In contrast, both PE-TSFM-3M4P and PE-TSFM-25M16P achieve substantial improvements in classifying degradation level 0, thereby enhancing the accurate identification of early degradation.

\begin{figure*}
    \centering
    \includegraphics[width=165mm]{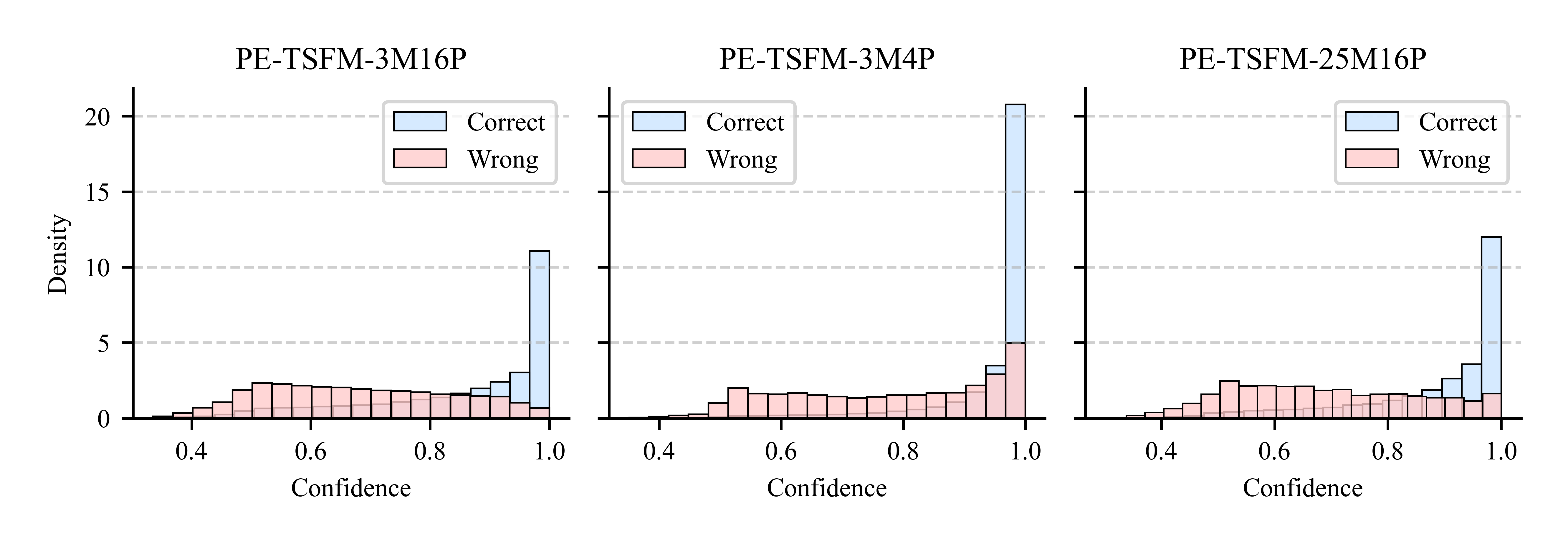}
    \vspace{-7mm}
    \caption{The confidence distributions of PE-TSFM-3M16P, PE-TSFM-3M4P, and PE-TSFM-25M16P. A larger probability density near 1 indicates a smaller uncertainty in model prediction, and a larger probability density near 0.5 indicates a greater uncertainty in model prediction.}
    \vspace{-5mm}
    \label{fig:mpp}
\end{figure*}

\subsection{Uncertainty and Confidence Analysis}

To further examine the robustness of the proposed PE-TSFM, we perform an uncertainty analysis based on its predictive probability outputs. Specifically, we use the maximum predicted probability (MPP) as a simple yet widely adopted confidence measure. The MPP is defined as the highest softmax probability among all candidate classes, representing the model’s most confident prediction. Generally, samples with higher MPP values are regarded as more confident results, while lower values indicate greater uncertainty. By comparing the distribution of MPP for correctly and incorrectly classified instances, we aim to evaluate whether the model’s confidence is well calibrated with its predictive performance.

Fig.~\ref{fig:mpp} presents the confidence distributions of the three PE-TSFMs on the entire test set. For all models, correctly classified samples exhibit confidence scores close to 1, whereas misclassified samples show much lower scores, suggesting that the predictive probabilities are more robust. Moreover, models with higher classification accuracy also display higher confidence scores for correct predictions. Among the three variants, PE-TSFM-3M4P achieves the highest confidence for correct predictions but also higher confidence for incorrect ones, suggesting potential overfitting to specific signal patterns. Conversely, PE-TSFM-25M16P maintains balanced confidence across classes, implying more stable calibration and improved generalization to unseen data.

These results indicate that increasing the model size is a more robust scaling way than decreasing the patch size, considering the model confidence.

\subsection{Feature Attribution and Interpretability Analysis}

\begin{figure}
    \centering
    \includegraphics[width=85mm]{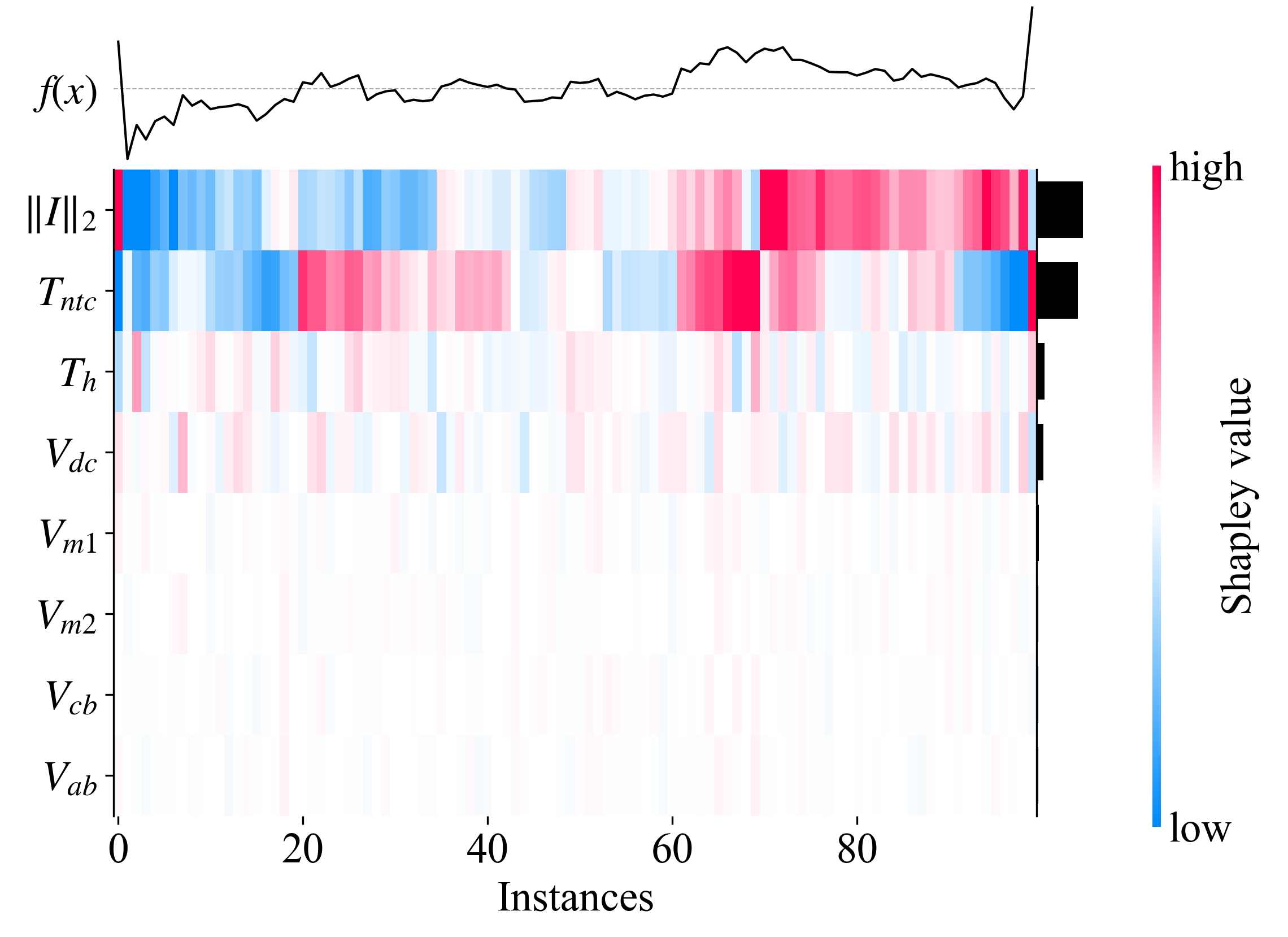}
    \vspace{-3mm}
    \caption{The Shapley values of each feature within the dataset.}
    \vspace{-5mm}
    \label{fig:shapley}
\end{figure}

To interpret the contribution of each input feature in the proposed multivariate time-series classification model, we employ the SHapley Additive exPlanations (SHAP) framework \cite{lundberg2017unified}. SHAP is a unified approach to feature attribution based on Shapley values from cooperative game theory, which measures the marginal contribution of each feature to the model prediction. Specifically, given a model $f$ and an input instance $x\in \mathbb{R}^{C\times L}$, the prediction in multi-variate time series classification tasks can be decomposed as $f(x)=\phi_0+\sum_{i=1}^C\sum_{j=1}^L \phi_{ij}$, 
where $C$ is the number of channels, $L$ is the sequence length, $\phi_0$ denotes the expected model output over the dataset, and $\phi_{ij}$ represents the attribution value (SHAP value) assigned to feature $i$ at the $j$-th step. A larger magnitude of $\phi_{ij}$ indicates a stronger influence of feature $i$ at the $j$-th step on the model’s output.

In this work, we utilize the GradientExplainer implementation of SHAP, which is based on the integrated gradient and can provide efficient approximate estimates of SHAP values for deep neural networks. Since our model operates on multivariate time series and outputs class probabilities, the raw SHAP values are computed across both the temporal dimension and the class dimension. To obtain a concise measure of global feature importance, we average the SHAP values over all time steps and predicted classes:
\begin{equation}
    \phi_i = \frac{1}{L\times O}\sum_{k=1}^O\sum_{j=1}^L |\phi_{ijk}|,
\end{equation}
where $O$ is the number of classes (4 degradation levels here) and $\phi_{ijk}$ means the SHAP attributions of feature $i$ at time step $j$ for class $k$. This aggregation allows us to compare the overall importance of different input variables, independent of temporal fluctuations or specific class outputs.

The SHAP values of 100 random test samples computed by PE-TSFM-3M16P are illustrated in Fig.~\ref{fig:shapley}, which shows that the most important feature for classification is the ac current $\|I\|_2$, the NTC temperature $T_{ntc}$, and the heat sink temperature $T_h$. This observation aligns closely with the physical degradation mechanisms of power modules, where increased on-state voltage and thermal resistance of power modules.

In summary, the SHAP analysis confirms that PE-TSFM not only achieves accurate classification but also produces physically interpretable feature attributions consistent with known degradation modes.

\section{Conclusion}

This paper presented a Power Electronics–tailored Time-Series Foundation Model (PE-TSFM) for noninvasive health monitoring of power converters, shifting from narrowly task-specific AI models to unified representation learning and unlocking the value of unlabeled operational data. The proposed framework fully relies on domain-specific pre-training and fine-tuning using PE data, enabling architecture customization for this application. A dual-attention mechanism was introduced to jointly capture temporal dependencies and inter-sensor physical correlations. Under unseen operating conditions, traditional time-series models achieved only 35–45\% accuracy, and fine-tuned generic TSFMs reached about 60\%, whereas the proposed PE-TSFM achieved approximately 92\% accuracy with far fewer parameters.

Ablation studies confirmed that the channel attention mechanism is essential; its removal caused an accuracy drop of up to 30\%. Although pre-training requires high-performance GPUs, the fine-tuned model has potential for edge implementation. Furthermore, both model size and patch size followed a scaling law in performance improvement; however, increasing model parameters provided more stable gains, whereas reducing patch size risked overfitting.

Finally, interpretability analysis revealed that current and temperature signals are the most influential features, aligning well with the physical degradation mechanisms of power converters.


\appendix
\textbf{Training for PE-TSFM:} The PE-TSFM is first pre-trained on the large-scale unlabeled dataset using masked reconstruction learning over 10 epochs, with a masking ratio of $30\%$. 
A warm-up cosine learning rate schedule is used, starting at $1\times10^{-6}$, peaking at $2\times 10^{-4}$, and decaying to $1\times10^{-7}$. 
Fine-tuning is then performed for degradation classification with a fixed learning rate of $1\times10^{-5}$, batch size of 128, early stopping (80\% for training and 20\% for validation) to prevent overfitting, and gradient clipping. 

\textbf{Training for Traditional Data-Driven Models:} The five baseline models are implemented using the Python library tsai with default configurations, open-sourced at GitHub
. The batch size for all methods is 128, and each deep learning-based model is trained for 100 epochs, with one cycle learning rate strategy peaking at $1\times10^{-3}$. 80\% of the labeled data in the training set for training, and the other 20\% for in-distribution performance evaluation, while the test set is for OOD performance evaluation.

\textbf{Training for the Other TSFMs:} The Moment
and Mantis
have been pre-trained on large-scale cross-domain datasets and open-sourced at Hugging Face. For fine-tuning, Moment follows the same configuration as PE-TSFM, except that the learning rate is set to $1\times10^{-4}$. Mantis is fine-tuned using a cosine learning rate schedule with warm-up, where the peak learning rate reaches $2\times 10^{-4}$. The model is trained for 100 epochs with a batch size of 128. Follow their recommendation, before feeding the data into Mantis, a dimension-reduction adapter based on Singular Value Decomposition (SVD) is applied to compress the multi-channel inputs into a single channel.

\printbibliography
\vspace{-10mm}
\begin{IEEEbiography}
[{\includegraphics[width=1in,height=1.25in,clip,keepaspectratio]{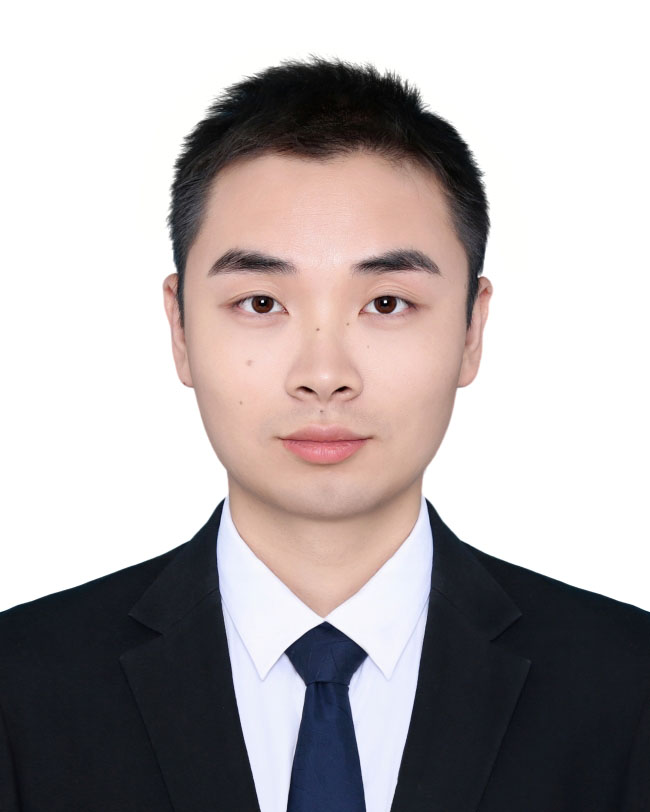}}]
{Xinyuan Liao} (Student Member, IEEE) received a B.Eng. degree in Computer Science and Technology from Ocean University of China in 2022 and an M.S. degree in Information and Communication Engineering from Northwestern Polytechnical University in 2025. He is currently working toward a Ph.D. degree at the Hong Kong Polytechnic University, Hong Kong. His current research interests include time-series foundation models and condition monitoring.
\end{IEEEbiography}

\vspace{-15mm}
\begin{IEEEbiography}[{\includegraphics[width=1in,height=1.25in,clip,keepaspectratio]{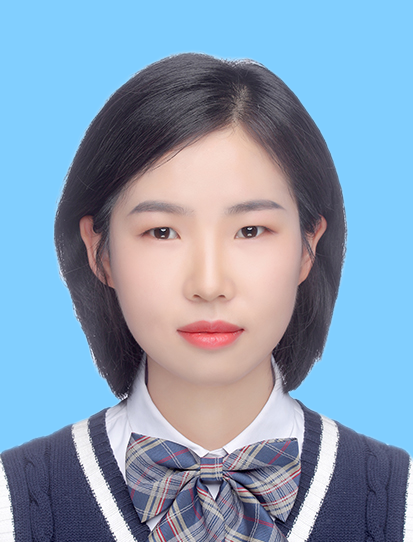}}]
{Xinyue Zhang} (Member, IEEE) received the B.S. and M.S. degrees in 2016 and 2019, respectively, and the Ph.D degree in 2025, all in electrical engineering from Northwestern Polytechnical University, China. She was a visiting student at Aalborg University, Denmark, from 2022 to 2023. She is currently working as a postdoctoral fellow with the Hong Kong Polytechnic University, Hong Kong. Her research interests include reliability and design of power electronics.
\end{IEEEbiography}

\vspace{-15mm}
\begin{IEEEbiography}[{\includegraphics[width=1in,height=1.25in,clip,keepaspectratio]{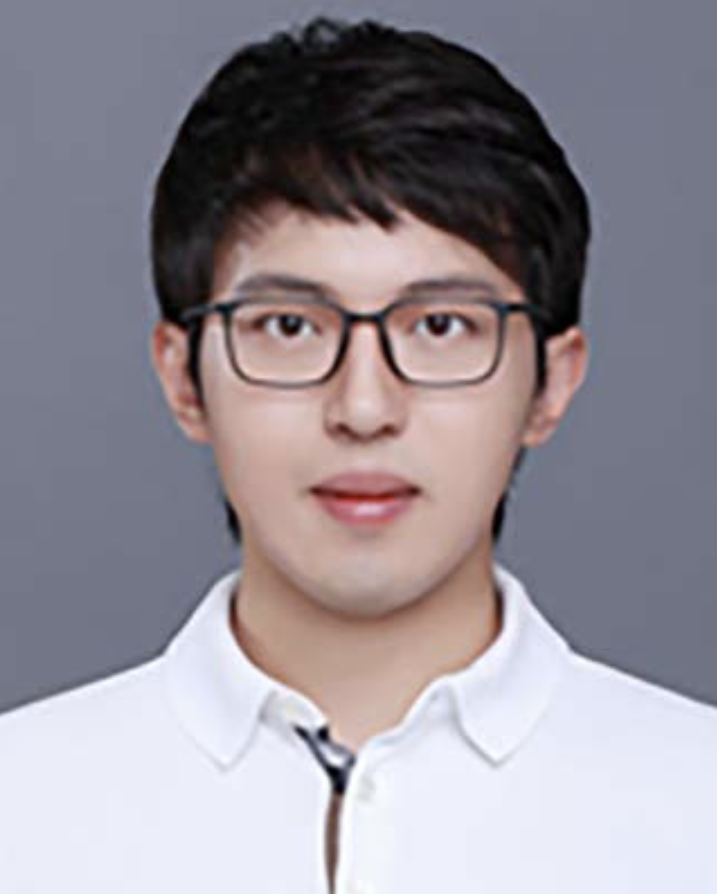}}]
{Xing Wei} (Member, IEEE) received the B.Eng. degree in electrical engineering and automation from Nanjing Normal University in 2016, the M.E. degree in electrical engineering from Southeast University in 2019, and the Ph.D. degree in power electronics from Aalborg University in 2024. He is currently a Postdoctoral Researcher with the Aalborg University. His research interests include power electronic reliability and condition monitoring.
\end{IEEEbiography}

\vspace{-15mm}
\begin{IEEEbiography}[{\includegraphics[width=1in,height=1.25in,clip,keepaspectratio]{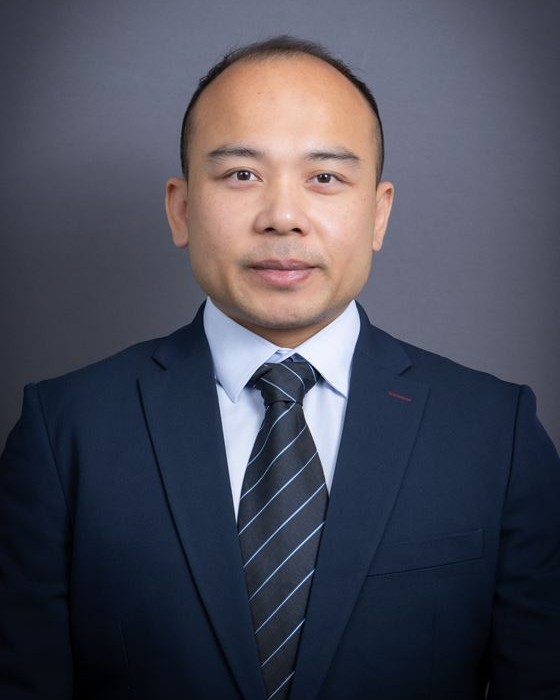}}]
{Junwei Liu} (Member, IEEE) received the B.Eng. degree in water conservancy and hydropower engineering from the Huazhong University of Science and Technology in 2012, and the Ph.D. degree in power electronics from the Hong Kong Polytechnic University in 2018. He is currently a Research Assistant Professor with the Hong Kong Polytechnic University, Hong Kong. His research interests include power electronics, wireless power transfer, transportation electrification, and renewable energy systems.
\end{IEEEbiography}

\vspace{-5mm}
\begin{IEEEbiography}[{\includegraphics[width=1in,height=1.25in,clip,keepaspectratio]{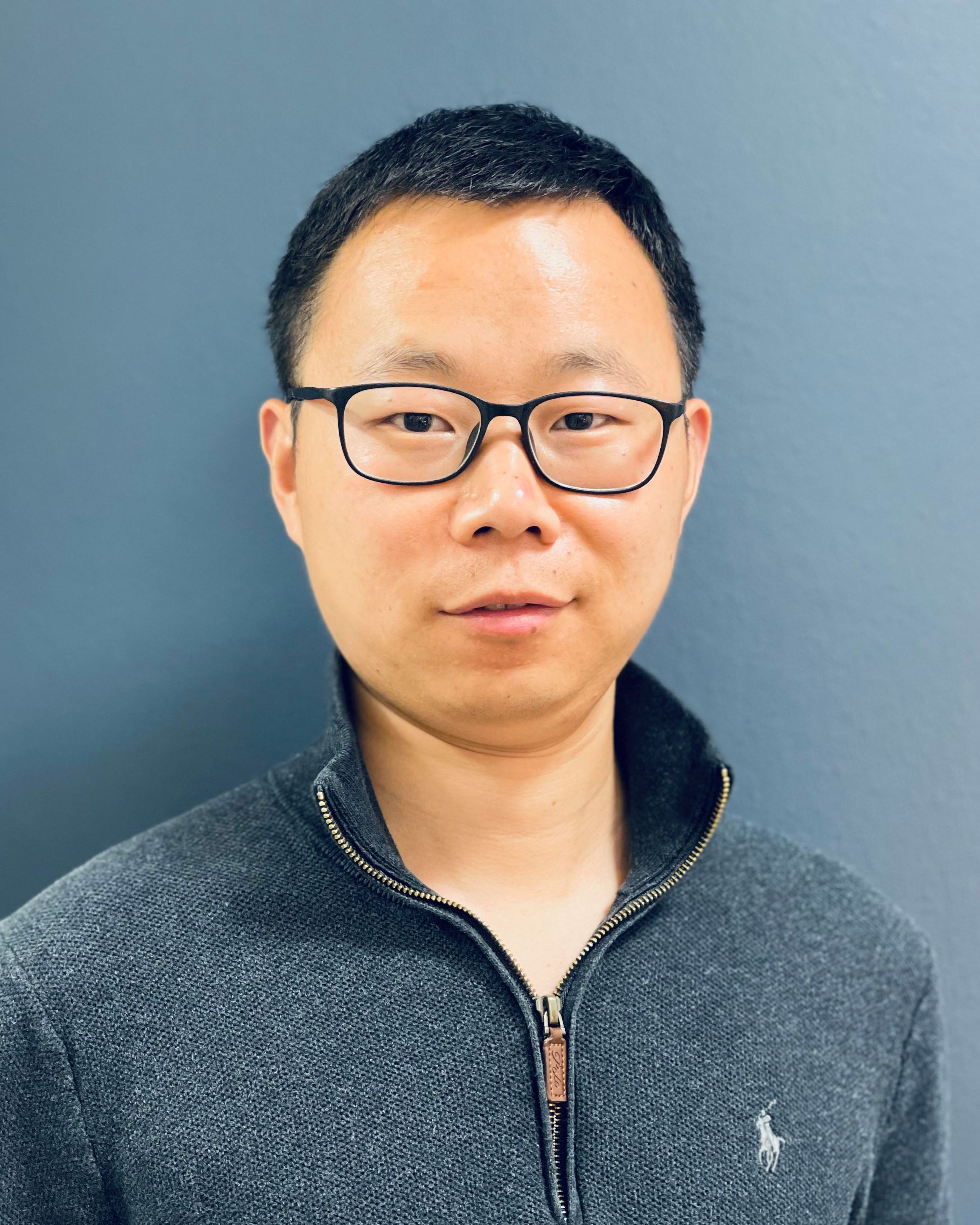}}]
{Shuai Zhao} (Senior Member, IEEE) received B.S., M.S., and Ph.D. degrees in Information and Telecommunication Engineering from Northwestern Polytechnical University, China, in 2011, 2014, and 2018, respectively. He is currently an Assistant Professor with AAU Energy, Aalborg University, Denmark. His research interests include physics-informed machine learning, system informatics, condition monitoring, diagnostics \& prognostics, and tailored AI tools for power electronic systems.
\end{IEEEbiography}

\vspace{-5mm}
\begin{IEEEbiography}[{\includegraphics[width=1in,height=1.25in,clip,keepaspectratio]{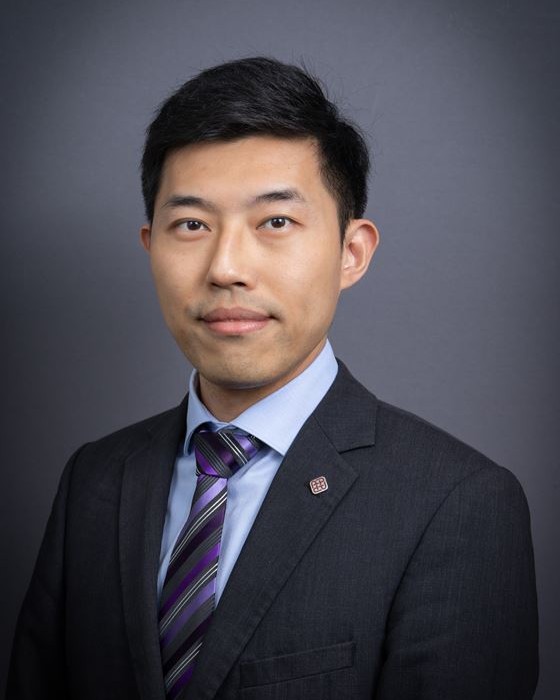}}]
{Siqi Bu} (Senior Member, IEEE) received the Ph.D. degree from the Electric Power and Energy Research Cluster, The Queen’s University of Belfast, Belfast, U.K., where he continued his postdoctoral research work before entering industry. He is currently a Full Professor and Associate Head with the Department of Electrical and Electronic Engineering, the Hong Kong Polytechnic University, Hong Kong. His research interests include power system stability, smart grid application, and transport electrification.
\end{IEEEbiography}

\vspace{-5mm}
\begin{IEEEbiography}[{\includegraphics[width=1in,height=1.25in,clip,keepaspectratio]{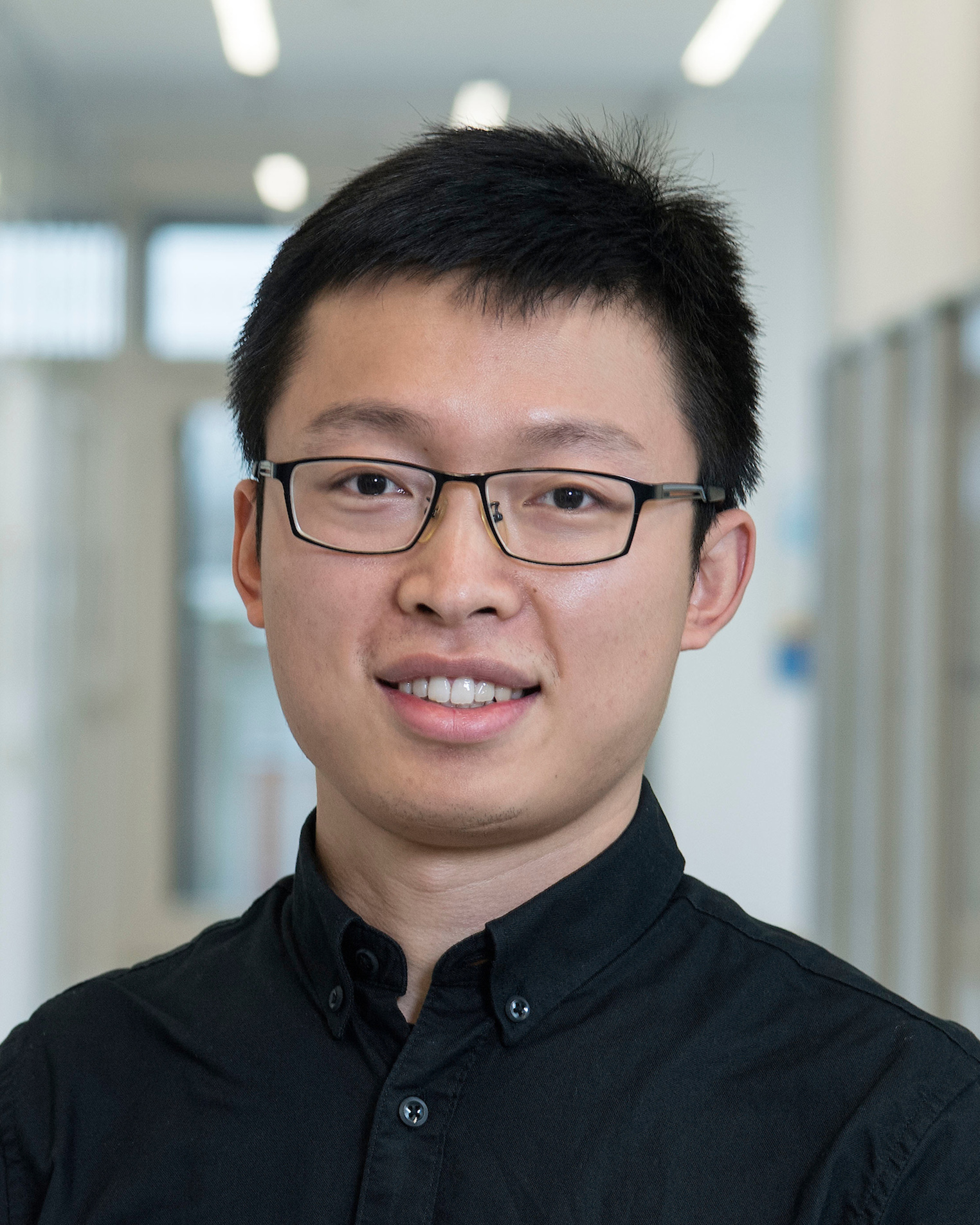}}]
{Yi Zhang} (Senior Member, IEEE) received the B.S. and M.S. degrees from the Harbin Institute of Technology, Harbin, China, in 2014 and 2016, respectively, and the Ph.D. degree from Aalborg University, Aalborg, Denmark, in 2020, all in electrical engineering. He is currently an Assistant Professor with the Hong Kong Polytechnic University, Hong Kong. His research focuses on advancing the reliability and sustainability of power electronics through design, testing, packaging, and condition monitoring.

Dr. Zhang was a recipient of the IEEE Power Electronics Society Ph.D. Thesis Award (2020), the First Place Prize Paper Award of the IEEE Transactions on Power Electronics (2021), the Second Place Prize Paper Award of the IEEE Transactions on Power Electronics (2025) and of the IEEE Transactions on Industry Applications (2025).
\end{IEEEbiography}

\vfill

\end{document}